# Search for Frame-Dragging-Like Signals Close to Spinning Superconductors

M. Tajmar, F. Plesescu, B. Seifert, R. Schnitzer, and I. Vasiljevich

*Space Propulsion and Advanced Concepts, Austrian Research Centers GmbH - ARC, A-2444 Seibersdorf, Austria*
*+43-50550-3142, martin.tajmar@arcs.ac.at*

**Abstract.** High-resolution accelerometer and laser gyroscope measurements were performed in the vicinity of spinning rings at cryogenic temperatures. After passing a critical temperature, which does not coincide with the material's superconducting temperature, the angular acceleration and angular velocity applied to the rotating ring could be seen on the sensors although they are mechanically de-coupled. A parity violation was observed for the laser gyroscope measurements such that the effect was greatly pronounced in the clockwise-direction only. The experiments seem to compare well with recent independent tests obtained by the Canterbury Ring Laser Group and the Gravity-Probe B satellite. All systematic effects analyzed so far are at least 3 orders of magnitude below the observed phenomenon. The available experimental data indicates that the fields scale similar to classical frame-dragging fields. A number of theories that predicted large frame-dragging fields around spinning superconductors can be ruled out by up to 4 orders of magnitude.

## INTRODUCTION

Gravity is the weakest of all four fundamental forces; its strength is astonishingly 40 orders of magnitude smaller compared to electromagnetism. Since Einstein's general relativity theory from 1915, we know that gravity is not only responsible for the attraction between masses but that it is also linked to a number of other effects such as bending of light or slowing down of clocks in the vicinity of large masses. One particularly interesting aspect of gravity is the so-called Thirring-Lense or Frame-Dragging effect: A rotating mass should drag space-time around it, affecting for example the orbit of satellites around the Earth. However, the effect is so small that it required the analysis of 11 years of LAGEOS satellite orbit data to confirm Einstein's prediction within ±10% (Ciufolini and Pavlis, 2004). Presently, NASA's Gravity-Probe B satellite is aiming at measuring the Thirring-Lense effect of the Earth to an accuracy better than 1% (Everitt, 2007). Therefore, apart from Newton's mass attraction, gravitational effects are believed to be only accessible via astronomy or satellite experiments but not in a laboratory environment.

That assumption was recently challenged (Tajmar and de Matos, 2003, Tajmar and de Matos, 2005, de Matos and Tajmar, 2005, Tajmar and de Matos, 2006a), proposing that a large frame-dragging field could be responsible for a reported anomaly of the Cooper-pair mass found in Niobium superconductors. A spinning superconductor produces a magnetic field as the Cooper-pairs lag behind the lattice within the penetration depth. This magnetic field, also called London moment, only depends on the angular velocity of the superconductor and the mass-to-charge ratio of the Cooper-pairs. By measuring precisely the angular velocity and the magnetic field, it is possible to derive the Cooper-pair mass with a very high precision, knowing that it consists of two elementary charges. The most accurate measurement to date was performed at Stanford in 1989 using a Niobium superconductor with a very surprising result: the Cooper-pair mass derived was larger by 84 ppm compared to twice the free-electron mass (Tate et al, 1989, Tate et al, 1990). The theoretical analysis predicted a Cooper-pair mass 8 ppm smaller than the free-electron mass including relativistic corrections. Several efforts were published in the last 15 years that investigated additional correction factors (Liu, 1998, Jiang and Liu, 2001). However, the discrepancy between measurement and theoretical prediction remained unsolved.

Within the classical framework, frame-dragging is independent of the state (normal or coherent) of the test mass. Over the last years, several theoretical approaches were developed that propose significantly amplified non-classical frame-dragging fields for superconductors with respect to normal matter (Tajmar and de Matos, 2006b, Chiao, 2007, Dröscher and Hauser, 2007, de Matos and Beck, 2007). Since 2003, an experimental program was established at the Austrian Research Centers (ARC) to search for such frame-dragging anomalies in the vicinity of spinning masses down to cryogenic temperatures using laser gyroscopes and accelerometers (Tajmar et al, 2006, Tajmar et al, 2007). This paper will give an overview of our experimental setup and the results obtained so far and will compare it with other recent tests performed by the Canterbury Ring Laser Group and the results from the Gravity Probe-B satellite.

## EXPERIMENTAL SETUP

The core of our setup is a rotating ring inside a large cryostat. Several rings have been used so far including Niobium and Aluminum to test a classical low-temperature superconductor as well as a non-superconductor material for reference purposes (outer diameter of 150 mm, wall thickness of 6 mm and a height of 15 mm), and a YBCO high-temperature superconductor (outer diameter of 160 mm and wall thickness of 15 mm). The ring can be rotated using a brushless servo motor or a pneumatic air motor to minimize any electromagnetic influence. According to the theoretical concepts, a frame-dragging-like field should be produced directly proportional to the superconductor's angular velocity. Another aspect of Einstein's theory is that a time-varying frame-dragging field should give rise to non-Newtonian gravitational fields, also called accelerational frame-dragging. Therefore, any angular acceleration of the superconductor should produce a gravitational field along the ring's surface. A short illustration of the expected fields around the rotating superconductor is shown in Fig. 1. Laser gyroscopes and low-noise accelerometers can be used to detect those frame-dragging fields if they are rigidly fixed to avoid any mechanical movement.

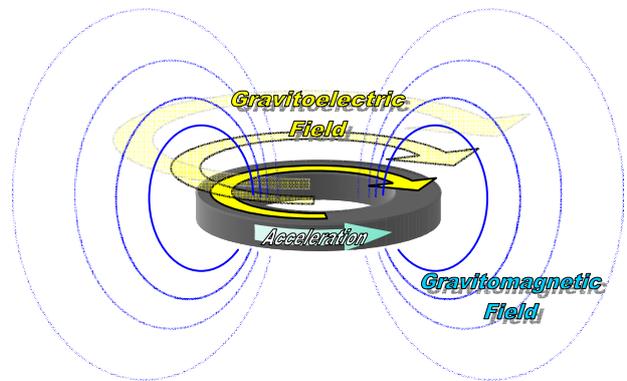

**FIGURE 1.** Gravitomagnetic and Gravitoelectric Field Generated by a Rotating and Angularly Accelerated Superconductor.

The experimental setup is illustrated in Fig. 2. The motor and the superconductor assembly are mounted on top and inside a liquid helium cryostat respectively, which is stabilized in a 1.5 t box of sand to damp mechanical vibrations induced from the rotating superconductor. The accelerometers and gyros are mounted inside an evacuated chamber made out of stainless steel, which acts as a Faraday cage and is directly connected by three solid shafts to a large structure made out of steel that is fixed to the building floor and the ceiling. The sensors inside this chamber are thermally isolated from the cryogenic environment due to the evacuation of the sensor chamber and additional MLI isolation covering the inside chamber walls. Only flexible tubes along the shafts and electric wires from the sensor chamber to the upper flange establish a weak mechanical link between the sensor chamber and the cryostat. This system enables a very good mechanical de-coupling of the cryostat with the rotating superconductor and the sensors even at high rotational speeds. A minimum distance of at least 5 mm is maintained between the sensor chamber and all rotating parts such as the motor axis or the rotating sample holder. In order to obtain a reliable temperature measurement, a calibrated silicon diode (DT-670B-SD from Lakeshore) was installed directly inside each rotating ring. A miniature collector ring on top of the motor shaft enabled the correct readout even during high speed rotation. Two temperature fixpoints enabled a temperature calibration during each run: the liquid helium temperature of 4.2 K and the evaluation of the critical temperature of the superconductor using the field coil. When the superconductor was cooled down, the field coil was switched on with a field below the critical field strength from the superconductors used. A Honeywell SS495A1 solid state Hall-sensor was installed inside the sensor chamber. Initially, the superconductor acted as a magnetic shield. But when the superconductor passed $T_c$, the magnetic field from the coil was recorded on the Hall sensor. At this point in time, the temperature read-out from the silicon diode must then correspond to the critical temperature.

The sample rings are glued inside an aluminum sample holder using STYCAST cryogenic epoxy. The bottom plate of the sample holder is made out of stainless steel as well as the rotating axis. In general, only non-magnetic materials were used throughout the facility, only the bearings are partly made out of steel with a non-negligible magnetic permeability.

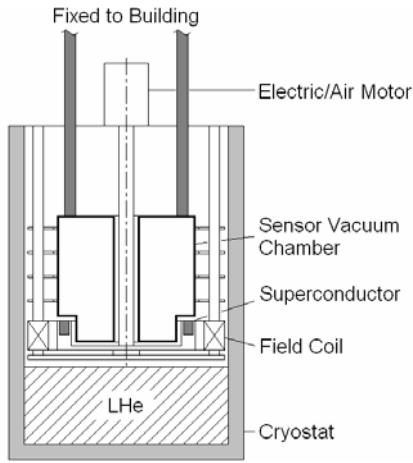 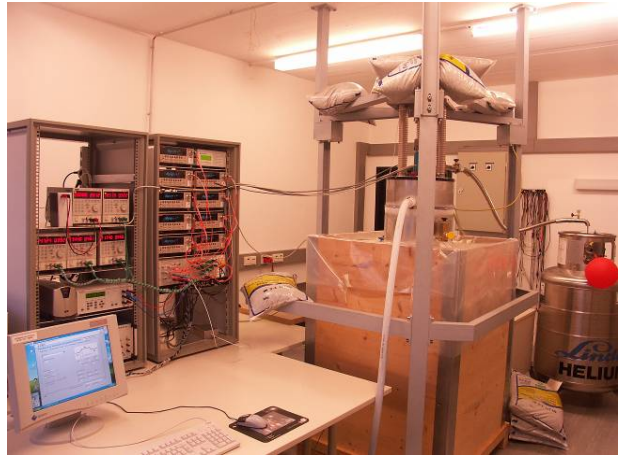

**(a)** Schematic Setup.     **(b)** Facility in the Lab at the Austrian Research Centers.

**FIGURE 2.** Experimental Setup.

The sensor-chamber can be equipped with accelerometers, gyros, temperature sensors and highly sensitive magnetic field sensors (based on the Honeywell HMC 1001 with 0.1 nT resolution). All sensors are mounted on the same rigid mechanical structure fixed to the upper flange. Each sensor level (In-Ring, Above-Ring and Reference) is temperature controlled to 25°C to obtain a high sensor bias stability. Trade-offs between different sensors can be found in (Tajmar et al, 2006).

## ACCELEROMETER MEASUREMENTS

After a survey on commercially available accelerometers, we decided to use the Colibrys Si-Flex SF1500S due to its low noise of 300 ng.Hz$^{-0.5}$, small size and low sensitivity to magnetic fields which we evaluated as $5\times10^{-4}$ g/T (expressing the acceleration in the unit of the Earth's standard acceleration). Using only the air motor, the magnetic fields inside the sensor chamber were always below 1 µT at maximum speed (originating from the rotating bearings). Therefore any magnetic influence is less than 1 ng and thus well below the sensor's noise level. The biggest systematic effect found is the so-called vibration rectification, a well known effect common to all MEMS accelerometers (Christel et al, 1991). Due to nonlinearities in the response of the pendulum, an anomalous DC offset appears when the sensor is exposed to vibration although the time average of the vibration of zero. Such vibrations are present due to the acoustic noise from the bearings as well as from the helium evaporation. Fortunately, the vibration rectification always leads to negative DC offsets independent of the angular speed orientation. By alternating between clockwise and counter-clockwise rotations and subtracting both signals, the anomalous DC offset can be eliminated and any real signal remains. A further improvement is to mount several sensors along the rings surface ("curl configuration") as shown in Fig. 3. This allows to further reduce sensor offsets and to increase the number of measurements. The accelerometer setup is described in more detail in (Tajmar et al, 2006). The accelerometers were read out using Keithley 2182 Nanovoltmeters with a measurement rate of 10 Hz.

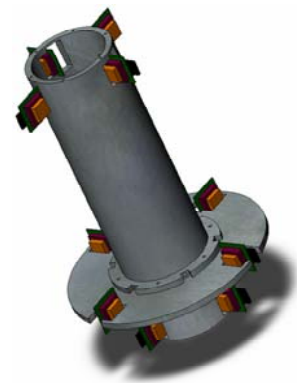

**FIGURE 3.** Accelerometer Insert for Sensor Chamber in Curl Configuration.

Many tests were carried out in the time frame from 2003-2006 to reduce the noise level on the sensors and to obtain an optimum mechanical de-coupling between the sensor chamber and the rotating parts of the facility (Tajmar et al, 2006, Tajmar et al, 2007). At the end we achieved a ground noise level on the sensors of a few µg$_{rms}$ while the ring was at rest and 20 µg$_{rms}$ when the ring was rotating. The noise level did not steadily increase with rotational speed but strongly increased above a speed of 350 rad.s$^{-1}$ (a resonance peak appeared at a speed of 400 rad.s$^{-1}$). In the final analysis, all sensor signals above this speed were therefore damped by a factor of 5. Using signal averaging over many profiles, the accuracy could be even further reduced.

The high accuracy tests were carried out using the Niobium ring. After the reduction of all vibration offsets, a signal above the noise level still remained when the ring was cooled down close to LHe temperature. An example is shown in Fig. 4 with the signal averaged plots from the in-ring position for the acceleration field and the applied angular acceleration to the ring. The difference between the temperature range in which the Niobium is superconducting ($g/\dot{\omega}$ = -2.26±0.3×10$^{-8}$ g.rad$^{-1}$.s$^2$) and normal conducting ($g/\dot{\omega}$ = -1.24±1×10$^{-9}$ g.rad$^{-1}$.s$^2$) is clearly visible. Also the correlation between measured acceleration and applied acceleration is good (0.78) but for the first peak only, the second sensor peak seems to precede the applied acceleration for 0.2 s and is less correlated. The signals were obtained in a differential configuration, i.e. the signals from the reference position were subtracted from the in-ring values. That is done to remove any mechanical artifacts, such as a real sensor chamber movement, and to reduce the noise level.

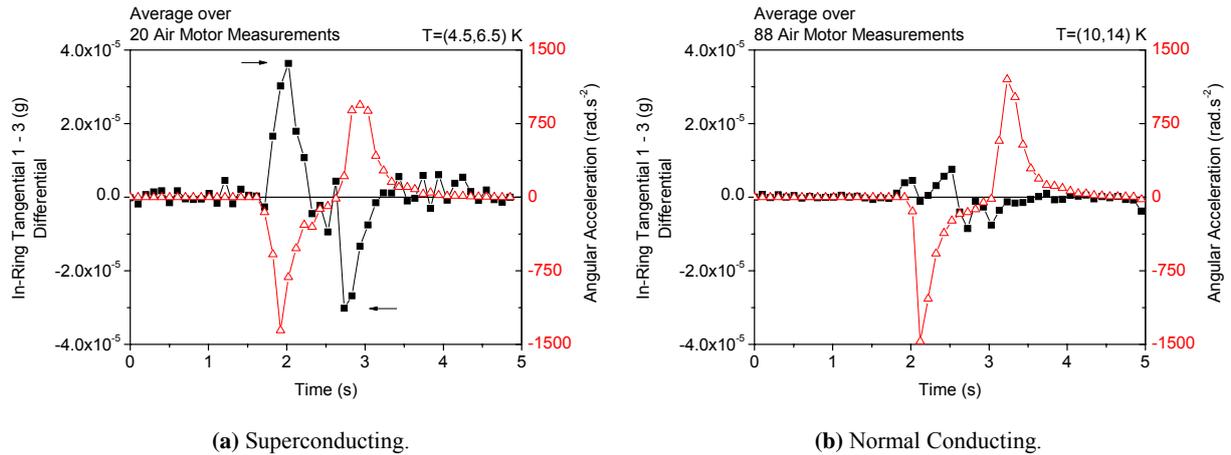

**(a)** Superconducting.    **(b)** Normal Conducting.

**FIGURE 4.** Signal Averaged In-Ring Sensor Data (■) Versus Applied Angular Acceleration (Δ).

If an angular acceleration of 1500 rad.s$^{-1}$ is applied to the superconducting ring, the tangential accelerometers show a counter-reaction of about 30 µg. What is the origin of these signals? The most important systematic error could still be the vibration rectification as also the helium evaporation is greatly contributing to the noise environment below 10 K. If the vibration sensitivity varies over the sensors, then our subtraction strategy in the curl configuration and also the subtraction of the reference position could maybe lead to false signals. Therefore, it was decided to further investigate this phenomenon using laser gyroscopes which are much less sensitive to vibration.

# LASER GYROSCOPE SETUP

## Gyro Measurements

Our requirements for the laser gyroscope include a low random angle walk (RAW), good bias stability and a high resolution, as well as a small size to fit in our sensor chamber. We therefore selected the KVH DSP-3000 fiber optic gyroscope that also features a digital output, which is much less affected by the electromagnetic environment compared to analog signals. It has a RAW of $2\times10^{-5}$ rad.s$^{-1}$ and a resolution of $1\times10^{-7}$ rad. Laser gyros are sensitive to magnetic fields due to the Faraday effect. We found values ranging from $2 - 0.5$ rad.s$^{-1}$.T$^{-1}$ depending on the gyro's axis. Since we measured maximum magnetic fields during rotation in the order of hundreds of nT, we decided to put each laser gyro into a μ-metal shielding box to further reduce magnetic influence. This reduced the maximum sensitivity to $0.04$ rad.s$^{-1}$.T$^{-1}$. At maximum speed, the magnetic influence is therefore below the gyro's resolution. However, due to the magnetic influence, the field coil must be off during the measurements as this can otherwise introduce sensor offsets when the superconducting ring passes $T_c$. This actually caused wrong signals in our first reported preliminary data (Tajmar et al, 2007). However, as these values were used to estimate the mechanical artifacts, the overestimated values gave a correct upper limit.

The gyro setup is illustrated in Fig. 5 showing the sensors inside the open vacuum chamber. Four gyros are mounted in three positions and two are mounted above the rotating ring (one showing up and one down to investigate any offset problems similar to the accelerometers). The axial distance from the top ring surface to the middle of each gyro position is 45.8 mm, 92.8 mm and 220.8 mm respectively. All sensors are mounted at a radial distance of 53.75 mm.

Due to the reduced vibration sensitivity, the following analysis was done without subtracting between alternating speed orientations or automatic subtraction between signals close to the spinning rings and the reference sensors. The gyro outputs for all three positions (reference, middle and above) using Niobium, Aluminum and YBCO samples with respect to the applied angular velocity of the rings is shown in Figs. 6 and 7. All profiles were recorded at an average temperature between 4 and 6 Kelvin. For reference purposes, we also removed the sample ring holder with the bottom plate but leaving the axis, bearings and motor assembly unchanged. The result is shown in Fig. 7b. The above position (LG 3-4) was obtained by subtracting the two above-ring gyros from each other due to their alternating orientations. That should eliminate any signal offsets if present. During each experimental test, more than 1000 profiles could be measured. Therefore, we were able to usually average over more than 20 profiles in every 2 K interval thus increasing statistical confidence.

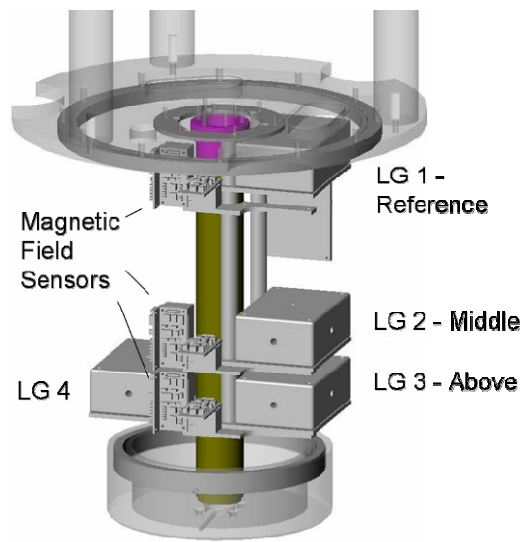

**FIGURE 5.** Laser Gyroscope Setup – Open Sensor Chamber (Outer Chamber Wall Removed for Illustration) and Rotating Ring with Sample Holder on Bottom.

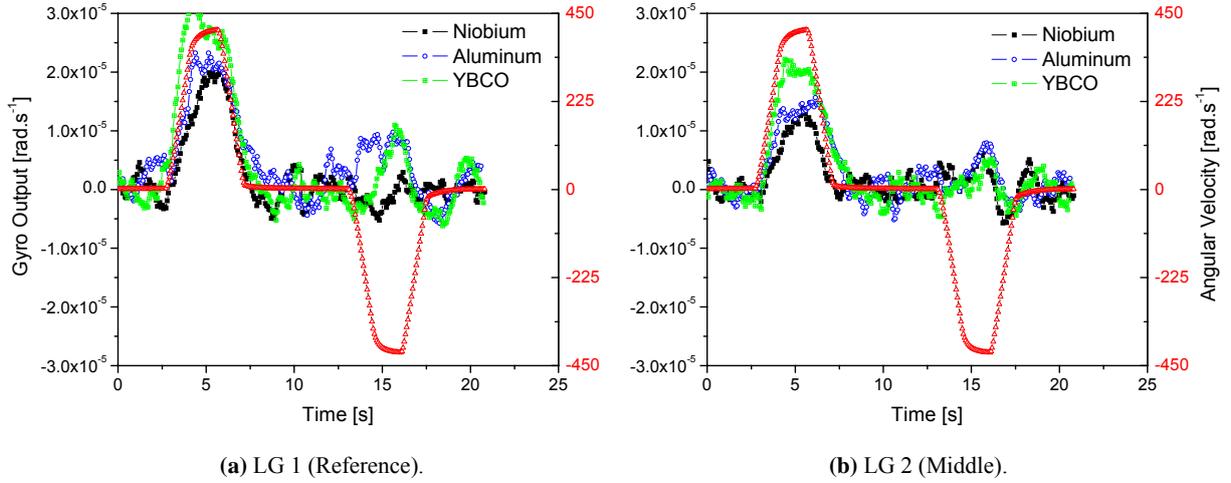

**(a)** LG 1 (Reference).  **(b)** LG 2 (Middle).

**FIGURE 6.** Laser Gyro Output for Niobium, Aluminum and YBCO Versus Applied Angular Velocity (Δ) between a Temperature of 4-6 Kelvin.

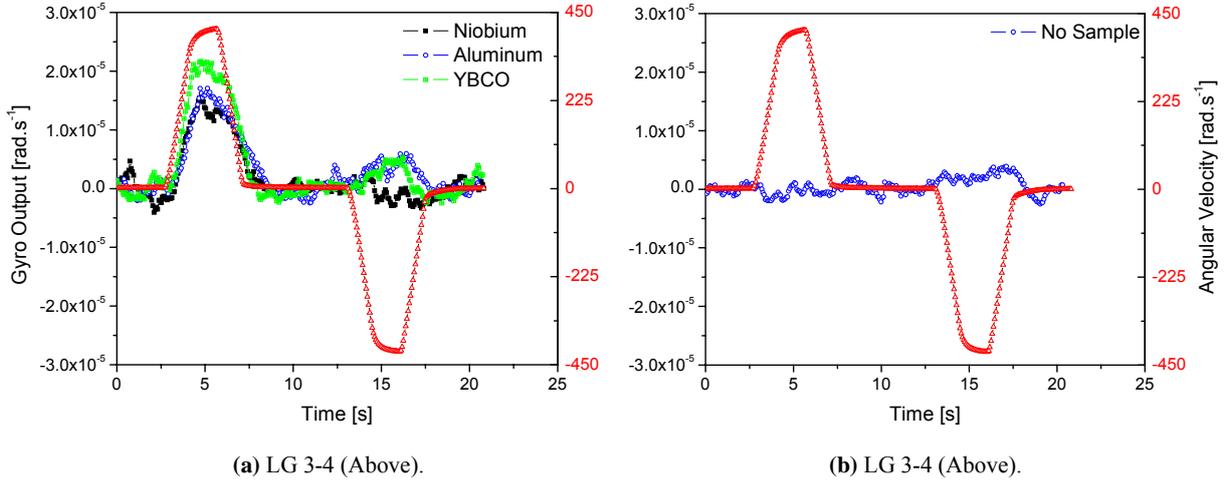

**(a)** LG 3-4 (Above).  **(b)** LG 3-4 (Above).

**FIGURE 7.** Laser Gyro Output for Niobium, Aluminum and YBCO (a) and without Sample Holder (b) versus Applied Angular Velocity (Δ) between a Temperature of 4-6 Kelvin.

The results are very surprising. First, the gyro outputs show a parity violation. Indeed, the gyro follows the applied angular velocity, but only if the ring is rotated in clockwise orientation. In order to check for any signal processing systematic, we cooled the YBCO sample down to 4.2 K and performed 40 successive clockwise rotations – and the gyro effect could always be measured. Then we cooled down again to 4.2 K and performed 40 successive counter-clockwise rotations – but any effect was an order of magnitude reduced compared to the clockwise rotation effect.

Second, YBCO gave the strongest signals while Niobium and Aluminum had similar responses to the applied angular velocity. Aluminum has a $T_c$ of 1.2 K and is therefore not superconductive at liquid helium temperatures. Therefore, the sample holder was made out of Aluminum as no signal contribution was expected from that material (the same applies to the stainless steel plate at the bottom). The difference of the YBCO sample with respect to the other ones is that the YBCO ring has a larger wall thickness (15 mm versus 6 mm) and that the outer diameter of the sample holder is also slightly larger (165 mm versus 160 mm). As the effect vanishes when the sample holder is removed, the observed gyro responses must be related to the sample holder and its sample rings.

Third, the effect does not decay as one would expect from a dipolar field distribution. The reference position in fact even gives the highest signal responses while the above and middle positions have similar values. Of course, we have to keep in mind that the laser gyro averages only the z-component of any present field over a large 89x48 mm area (more specifically, the fiber coil has an elliptic shape) and it is therefore impossible to derive a field distribution using these measurements. Nevertheless, one would have expected that the field decays over distance if the source is the spinning ring. It is important to note that for the case of the above-ring gyros, if the gyro's orientation is flipped, also the effect sign flips and the effect is only present for the same clockwise-rotating of the spinning ring. The measurement also tells us that the effect is at least rotationally symmetric.

And fourth, the gyro responses do not correlate with the accelerometer measurements if one assumes the standard induction law. Any parity violation was not visible as the accelerometer measurements were always done subtracting the alternating speed profiles from each other in order to eliminate the anomalous DC offsets from the vibration rectification. From the induction law, the Niobium signals should be higher by a factor of about 100. Assuming the validity of the gyroscope measurements, that means that either the accelerometer measurements are vibration artifacts or the standard induction law does not apply. That could be explained by the breakdown of the usual weak-field approximation used to derive the Maxwell-like equations out of general relativity. Also, new theoretical concepts actually predict that the acceleration-induced effect is stronger by about two orders of magnitude compared to the gyro response as observed in our measurements (de Matos and Beck, 2007). However, the measurements so far already rule out our initial theoretical approach that modeled the effect as proportional to the ratio between the matter densities in the coherent state with respect to the lattice (Tajmar and de Matos, 2006b).

The ratio of the above gyro versus angular velocity response for clockwise rotation with respect to temperature for Niobium, Aluminum and YBCO is shown in Fig. 8. It is interesting to see that the effect occurs below a critical temperature that does not coincide with the superconducting temperature of the materials. For Niobium and Aluminum, the critical temperature is about 16 K whereas for YBCO it is close to 32 K. Apart from the value at 4 K, the Niobium curve is above Aluminum by about $1\times10^{-8}$. The YBCO curve is quite constant below 32 K until the Aluminum critical temperature at 16 K. It seems like the effect of Aluminum adds then to the YBCO curve. So either Aluminum does indeed produce an effect and therefore contributes to Nb and YBCO due to the Al sample holder material, or Aluminum provides a reference case due to the vibration environment that leads to gyro sensor offsets. If the Aluminum part is removed from the curves, than YBCO has a more or less constant coupling factor of $2.2\times10^{-8}$ below 32 K and Niobium $1.6\times10^{-8}$ below 16 K.

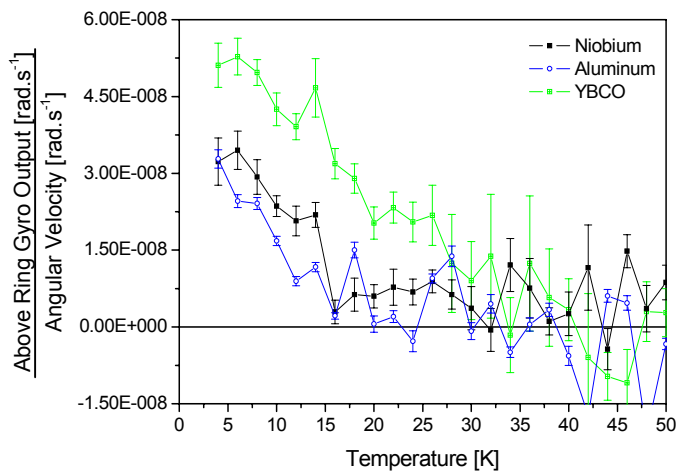

**FIGURE 8.** Variation of Above Ring Gyro Output versus Angular Velocity with Temperature (for Clockwise Rotation).

## Systematic Effects

*Electromagnetic Fields*

The sensors are inside an evacuated vacuum chamber, which is grounded and acts as a Faraday cage. The supply voltages for the gyros are generated and stabilized inside the vacuum chamber. Moreover, the gyro output is digital, which greatly eliminates possible electromagnetic interference of the signals along their transmission to the computer. From the magnetic sensitivity, an induced coupling factor of $4\times10^{-11}$ has been measured, which is 3 orders of magnitude below the observed effect. In addition, gyro measurements were done with and without μ-metal shielding of the gyros with very similar results. Therefore, the explanation of the observed effects due to electromagnetic effects is very unlikely.

## Pressure Effect

The rotation of the ring causes a strong evaporation of the liquid helium. This pressure increase can maybe tilt or turn the vacuum chamber and cause sensor offsets. We investigated the pressure increase of the helium gas in the cryostat by mounting a Keller PA-23 pressure transmitter on the top of the facility. After filling up of the facility with liquid helium, the first profile caused a peak pressure increase of 100 mbar as some liquid was evaporated due to the stirring of the rotating ring. All successive profiles showed no change in the pressure within the sensor resolution of 3 mbar. As a worst-case scenario, we simulated the pressure increase using compressed air that was connected to the liquid helium line at room temperature. Instead of running the motor, the pressure was increased during each profile until the sensor measured the 100 mbar. The result is shown in Fig. 9. No gyro response other than the usual noise is seen during the pressure increase. Therefore, the pressure increase due to liquid helium expansion cannot lead to the observed effects and especially no parity violation.

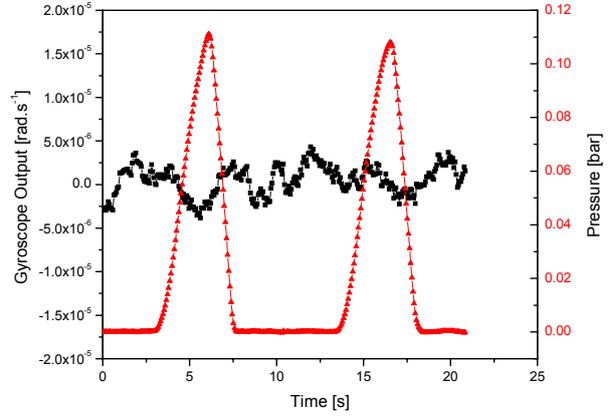

**FIGURE 9.** Gyro Output Versus Pressure.

## Vibration Offsets

Although the manufacturer noted no vibration sensitivity, we performed a dedicated test by putting the gyro on a table next to a shaker table. We were indeed able to produce a vibration offset but only using very large amplitudes at a frequency of 60 Hz, which is similar to the frequencies of the air motor at maximum speed as shown in Fig. 10. This offset was always negative and it was sensitive to the orientation of the sensor axis on the shaker table. We found out that the vibration offset was only present when the gyro's axis was pointing to the Z+ or X+ direction (Z+ gave the strongest response). The difference between the two directions is also evident in Fig. 10. Pointing the gyro's in the Y direction showed no vibration offsets other than noise. The magnitude of the vibration offset was directly related to the amplitude of the vibration. When the amplitude was reduced so that the noise of the gyro's output did not increase during the vibration, the vibration offset vanished below detectability ($< 2 \times 10^{-6}$ rad.s$^{-1}$).

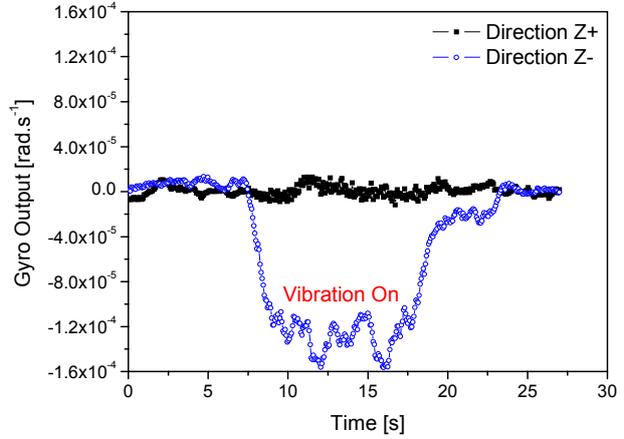

**FIGURE 10.** Gyro Output Versus Vibration.

During all gyro measurements, no increase in the noise level of the gyros was detected during the ring's rotation similar to the case as described above. The measured gyro's response to the ring's rotation was positive and not negative as the vibration offsets. The same signal with alternating sign was also measured by flipping the gyro's orientation axis as in the case of LG3 and LG4 which would not be possible with the vibration offsets. Also the difference between the YBCO and Nb/Al rings as well as the parity violation cannot be explained. Therefore, the explanation of our gyro results as vibration offsets is not likely – but it is the most important error source that has to be further analyzed since the real noise environment is more complex than the signals produced from the shaker table.

*Sensor Tilting*

Due to the helium gas flow, the sensor chamber might be tilted. Since the gyro measures the Earth's rotation, tilting of the sensors can induce a false signal due to the different offset from the Earth's measurement. Assuming a tilting angle α, the offset can be expressed as

$$\text{Offset} = \text{Offset}_{Earth} \cdot [\sin(\text{latitude}) - \sin(\text{latitude} + \alpha)] \quad . \tag{1}$$

Therefore, in order to get offsets in the gyro's signal similar to the observed effects, a tilting angle of at least 20° is necessary with $\text{Offset}_{Earth}=73\times10^{-6}$ rad.s$^{-1}$ and a latitude of 48°. It is impossible that the sensor chamber tilted by as much as 20° during each profile. This artifact can be therefore ruled out.

*Mechanical Friction*

During the rotation of the ring, there is a 5 mm gap with helium gas between the sample holder and the sensor vacuum chamber. Since the viscosity of helium gas at 5 K is an order of magnitude below the viscosity of air at room temperature (Keller, 1957), mechanical friction from the rotating helium gas cannot explain the observed effects as the effect only occurs when passing through a critical cryogenic temperature. Nevertheless, we shall examine the order of magnitude from such friction effects. We therefore built a finite element model of the sensor chamber with its connecting rods using ANSYS as shown in Fig. 11. The force on the bottom of the sensor chamber from the rotating gas can be calculated using Stoke's law as

$$F_{Stokes} \cong \eta \cdot \frac{Av}{\Delta x} \quad , \tag{2}$$

where $\eta$ is the viscosity, $A$ the area, $v$ the velocity of the gas and $\Delta x$ the gap between the rotating ring and the sensor chamber. Using our geometry we obtain a maximum friction force on the bottom of the sensor chamber of about 1 mN at maximum speed. Using the finite element model, this force leads to a drilling of the sensor chamber assembly of $5\times10^{-8}$ rad. The upper limit false coupling factor from friction is therefore $5\times10^{-11}$, which is three orders of magnitude below the observed effects similar to the magnetic sensitivity.

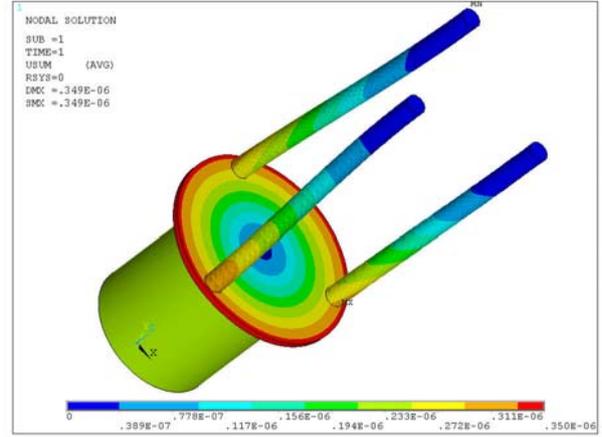

**FIGURE 11.** Finite-Element Analysis of Sensor Chamber and Connecting Rods Drilling.

## Discussion

The effects and the analysis so far lead to the following possible interpretations:

1. The effects are real: A frame-dragging-like signal was detected from spinning rings at cryogenic temperatures. The effect occurs below a critical temperature which does not coincide with the superconducting temperature of the rings. The strength of the effect depends on the material of the ring. The coupling factor of the observed effect with respect to the applied angular velocity is in the range of 3-5$\times10^{-8}$. We observed a parity violation, such that the effect is dominant in the clockwise rotation (when looking from above) in our laboratory setup. No classical or systematic explanation has been found so far for the observed effects. The field expansion is not clear at the moment and needs further investigation.

2. The reference signal has to be subtracted from the measurements due to sensor chamber movements: The coupling factor is then reduced to about $1.3\times10^{-8}$. Now it is less clear if the effect's origin could be indeed related to superconductivity because Nb still shows a signal but Al does not as shown in Fig. 12 (they have the same sample holder dimensions). Nevertheless, the critical temperature for the effect is different than the superconducting critical temperature. Also in this case, parity violation is observed.

3. Aluminum is the reference case which has to be subtracted from the measurements: The vibration environment from the helium gas expansion causes sensor offsets. Since aluminum is not a superconductor, this material can be considered the reference case which has to be subtracted from the results. As discussed above together with Fig. 8, the coupling factor for Nb is then reduced to $1.6 \times 10^{-8}$ and for YBCO to $2.2 \times 10^{-8}$. The effect is now indeed related to superconductivity, however as in the cases discussed above, the critical temperature does not coincide with the superconducting temperature. We still observe a parity violation.

4. All signals are false due to systematic effects: Of course this may still be a possibility. The strongest indication is that the signals do not decay over the three positions measured. As this looks like giving room to facility artifacts, it could be also possible that the field is propagating along the spinning motor axis. All systematic effects analyzed so far contribute to less than 3 orders of magnitude to the observed effects. Only vibration effects may still contribute to the gyro output but they seem unlikely to explain all different aspects of the effect such as parity violation and the dependence on the ring material.

Since the systematic effects analyzed so far are below the laser gyro measurement accuracy and therefore cannot account for our measurements, we suggest that the first interpretation is correct. That in turn puts severe limits on theoretical models that were proposed to predict frame-dragging fields generated by superconductors as the phenomenon that we observe is apparently not related to superconductivity and shows a parity violation.

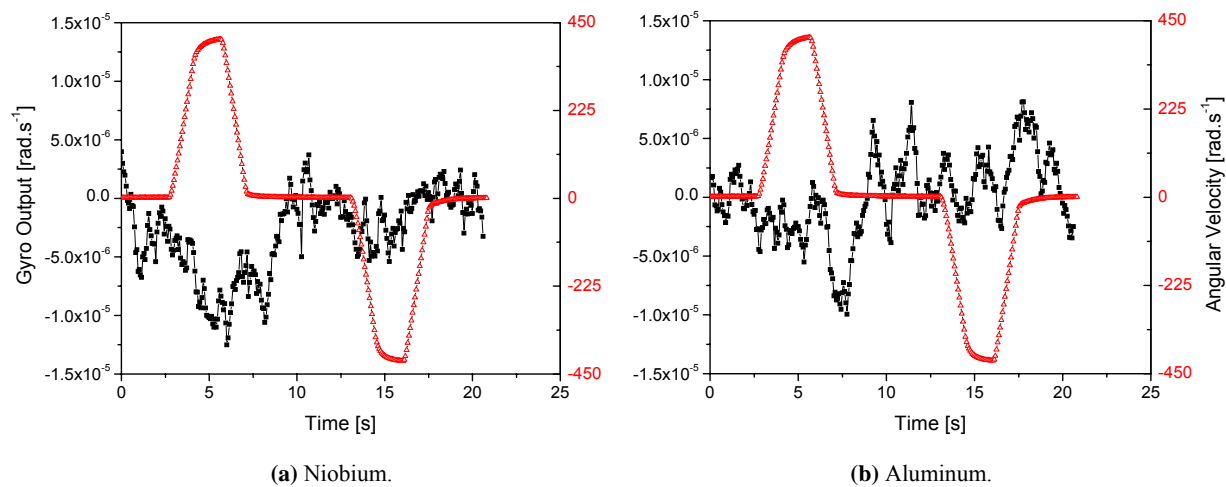

(a) Niobium.    (b) Aluminum.

**FIGURE 12.** Above Laser Gyro Output Minus Reference (LG3-4 – LG1) Versus Applied Angular Velocity (Δ) between a Temperature of 4-6 Kelvin.

## COMPARISON WITH OTHER EXPERIMENTS

In order to determine if our results are indeed genuine or facility artifacts, it is important to compare them with other experiments and to check for consistency. Fortunately, at least two similar experiments are available that can be used for such a comparison: a recent test of a spinning lead superconductor close to the world's largest ring laser gyroscope (Canterbury Ring Laser Group) and the Gravity-Probe B satellite using spinning superconducting gyroscopes to detect Earth's frame dragging field. Due to differences between the experiments, models are necessary for such a comparison to extrapolate our results to the other setups. In addition to the different models that were already proposed (Tajmar and de Matos, 2006b, Dröscher and Hauser, 2007, de Matos and Beck, 2007), we will follow a phenomenological approach assuming that the fields detected are an amplification of classical frame-dragging fields without linking it to superconductivity. After passing a critical temperature, the classical frame-dragging fields $B_{g0}$ are enhanced using the simple expression

$$B_g = \gamma \cdot B_{g0} \quad . \tag{3}$$

According the our measurements so far, the enhancement factor γ (or frame-dragging relative permeability factor) is temperature and material dependent, and shows a parity violation. For the comparison with our experiments and the Canterbury setup, the classical frame-dragging field $B_{g0}$ at the center of a ring and a disc is given by,

$$B_{g0} = \frac{4G}{c^2} \frac{m}{R_o + R_i} \cdot \omega \quad , \tag{4}$$

where $m$ is the spinning mass, $R_0$ the outer radius, $R_i$ the inner radius (=zero for the case of a disc), and $\omega$ the angular velocity. Using our measurements, we get $\gamma \cong 1.2 - 1.7 \times 10^{18}$ for the various material combinations at a temperature of 5 K (we assumed that the field measured at the gyro's location is similar to the one in the center of the spinning ring).

For the comparison with the Gravity-Probe B data, we have to calculate the classical frame-dragging field for a spinning shell or sphere with radius $R$ along the central axis $z$, given by

$$\begin{aligned} B_{g0}(z) &= \frac{2G}{c^2} \frac{I}{z^3} \cdot \omega \quad z > R \\ B_{g0}(z) &= \frac{5G}{c^2} \frac{I}{R^3} \cdot \omega \quad z = 0 \quad (sphere) \\ B_{g0}(z) &= \frac{2G}{c^2} \frac{I}{R^3} \cdot \omega \quad z = 0 \quad (shell) \end{aligned} \tag{5}$$

where $I$ is the moment of inertia.

It is important to note that the classical fields scale with mass and geometry of the spinning source. If the effect would be related to a superconductive-like phenomenon, the field strengths are expected to be independent of mass and geometry of the spinning source similar to the London moment-magnetic field produced by a rotating superconductor that is only proportional to the angular velocity and the charge-to-mass ratio of the Cooper-pair.

## Comparison with Canterbury Ring Laser Group Experiment

Recently, an independent experimental test was carried out, where a lead disc at liquid helium temperature was spinning close to the world's most precise ring laser gyro UG2 from the Canterbury Ring Laser Group (Graham et al, 2007). Contrary to our setup, the gyro here is operated outside of the cryostat facility due to its large dimensions of 21x39.7 m. This should greatly reduce any vibration offsets associated with the evaporating helium gas. Fig. 13a shows the gyro's response to the speed of the spinning superconductor[*]. Here too, we see that the gyro reacts if the superconductor is rotated. Again, we note a parity violation as the gyro response is greater for the counter-clockwise rotation – which is the opposite direction as in our experiments. Since this experiment was carried out in the southern hemisphere and our experiments in the northern hemisphere, a first hint at the origin of the parity effect could be the Earth's rotation. A similar parity anomaly that may be related to our effect was reported on gyro weight and free-fall experiments in a Japanese laboratory (Hayasaka and Takeuchi, 1989, Hayasaka et al, 1997), which showed the same parity direction as in our laboratory and was also located on the northern hemisphere. However, these claims could not be verified in a number of replication attempts up to now (Luo et al, 2002 and references therein).

As the UG2 gyro is large compared to the actual rotating superconductor, a field distribution has to be assumed in order to compare our results with the UG2 results. The Canterbury group used a dipolar distribution for the interpretation of their experimental data. As computed in (Graham et al, 2007), the UG2 gyro output was multiplied by a factor of $1.7\times10^6$ to get an estimate of the frame-dragging-like signal strength in the vicinity of the rotating superconductor. Fig. 14b shows the gyro response corrected by the dipolar field distribution and the applied angular velocities. The coupling factor (Gyro Output)/ω computed for the counter-clockwise direction is $3.8\pm3\times10^{-7}$. It is possible to improve the statistics with additional filter. By applying a 200 pt digital moving average filter, the counter clockwise direction coupling factor yields $3.8\pm1.3\times10^{-7}$. This is nearly one order of magnitude above our measurements in the close vicinity of the spinning superconductor. One major difference is the disc shape in comparison with our ring superconductors.

---

[*] The gyro data from (Graham et al, 2007) was converted into rad.s$^{-1}$ and the offset at zero angular velocity was removed for consistency with the present paper.

Indeed, by computing the enhancement factor γ from the angular momentum of the spinning disc and the UG2 gyro output, we get $\gamma \cong 1.9\times10^{18}$ for the counter-clockwise rotation, a value very close to the ones in our setup. This suggests that the fields observed are scaling like classical frame-dragging fields. A superconducting-like phenomenon such as a gravitomagnetic London moment would give the same results in the Canterbury and in our setup.

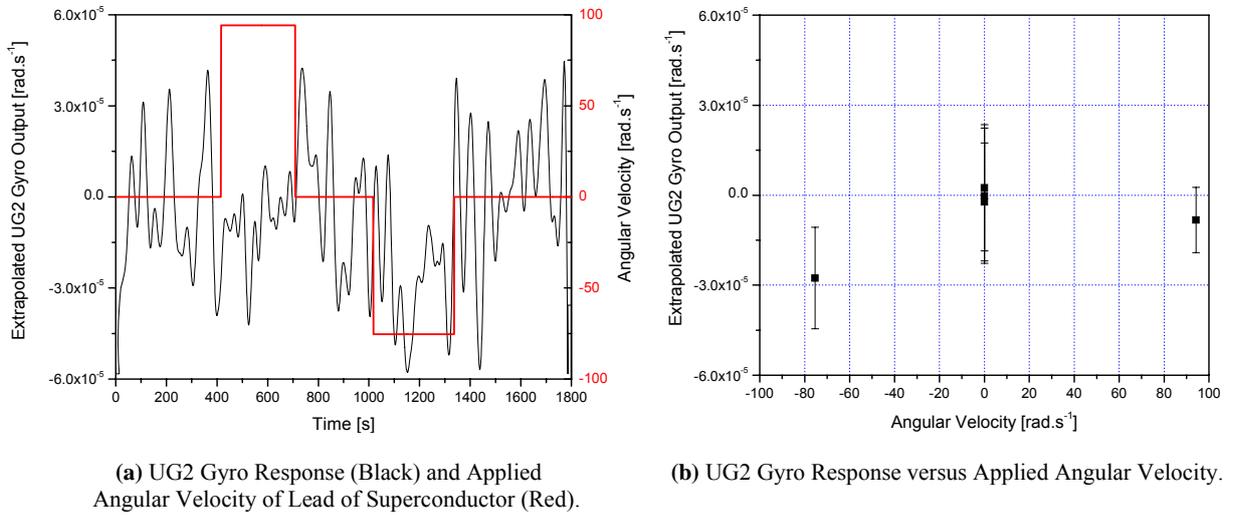

(a) UG2 Gyro Response (Black) and Applied Angular Velocity of Lead of Superconductor (Red).

(b) UG2 Gyro Response versus Applied Angular Velocity.

**FIGURE 13.** Experimental Results from the Response of the UG2 Ring Laser Gyro to the Rotation of a Lead Superconductor – Extrapolated to the Spinning Source and Normalized (Data taken from Graham et al, 2007).

## Comparison with Gravity-Probe B Results

Since Gravity-Probe B (GP-B) measured with unprecedented accuracy the precession of superconducting gyros in a polar orbit around Earth, one would expect to see the effect of an enhanced frame-dragging-like field in their setup as well. The experiment consists of four gyros, which are equally spaced and aligned along the roll axis of the satellite towards the guide star IM Pegasi (Keiser et al, 1998). Gyro 1 and 3 are rotating in one direction and Gyro 2 and 4 in the other direction so that the poles are always facing each other. Note that for the gravitational case, similar poles attract contrary to magnetic fields where opposite poles attract (Wald, 1972). If the spinning gyros now produce a frame-dragging-like field, a restoring torque would appear proportional to the misalignment of the gyro's axis with the spacecraft axis towards the guide star. This in turn will cause additional precession of the gyros. We can express this drift in function of the misalignment angle $\psi$ as (Forward, 1961, Muhlfelder, 2007)

$$\Omega = \frac{B_g}{2} \cdot \sin(\psi) \quad , \tag{6}$$

where $B_g$ is the frame-dragging-like field at the location from one gyro caused by the others. As a first approximation, we use a dipolar field expansion with a gyro separation of 75 mm and a gyro diameter of 38 mm.

The final speed of the four gyros was measured to be 79.4, 61.8, 82.1 and 64.9 Hz respectively. The gyros are made out of fused quartz spheres coated with a 1.25 μm layer of Niobium.

An anomalous torque proportional to the misalignment angle was indeed seen in the GP-B experiment with drift rates of a couple of arcsec/day/degree. The anomalous torque anomaly is presently modeled as an electrostatic patch effect due to a variation of the electric potential along the gyro's surface. Without distinguishing between a patch effect or frame-dragging origin of this effect, we can at least express an upper value for any non-classical frame-dragging field generated by the rotating superconducting Nb shells. Using the average torque of all four gyros, the upper-limit coupling factor $B_g/\omega \cong 1\times10^{-9}$ at the center of the spinning superconductor. This is more than an order of magnitude smaller compared to our setup.

By using the angular momentum of the gyro, we can estimate an upper limit for the enhancement factor by comparing the fields to the observed anomalous torques. Since the gyro has only a Nb layer, the $SiO_2$ will dominate the angular momentum by orders of magnitude. By taking an enhancement factor γ for $SiO_2$ that is half the value for Nb, we get a fairly good match with the observed anomalous torques as shown in Fig. 14 (here we assumed again a parity violation similar to the other experiments). This suggests that our effect might be an alternative explanation for the anomalous torques observed on Gravity-Probe B. Since we did not specifically measure $SiO_2$ in our setup, we can set this enhancement factor as a possible upper limit for this material at 4 K. Since Nb contributes very little to the angular momentum, the upper limit enhancement factor is about a factor of 500 larger compared to the values from our experiment. That would still leave enough room for a possible patch effect assuming that $SiO_2$ does not contribute to the frame-dragging effect. Further experiments are necessary to clarify this point.

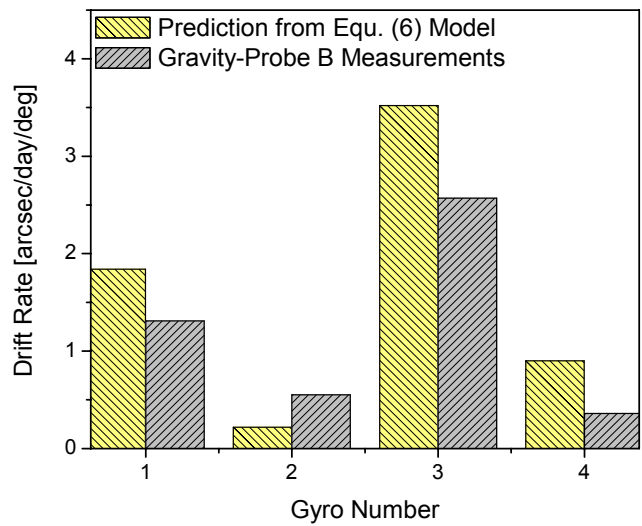

**FIGURE 14.** GP-B Gyro Misalignment Torque Drifts and Spin-Spin Approximation from Equ. (6) - Data taken from (Gill and Buchman, 2007) with Linear Fit up to 1 deg and $\gamma_{SiO2}=0.6\times10^{18}$.

# SUMMARY AND CONCLUSION

High-resolution accelerometer and laser gyroscope measurements were performed in the vicinity of spinning rings at cryogenic temperatures. After passing a critical temperature, which does not coincide with the material's superconducting temperature, the angular acceleration and the angular velocity applied to the rotating ring could be seen on the sensors although they are mechanically de-coupled. A parity violation was observed for the laser gyroscope measurements such that the effect was greatly pronounced in the clockwise-direction only.

Table 1 summarizes our laser gyroscope experiments and compares them with an independent test performed by the Canterbury Ring Laser Group and the Gravity-Probe B anomalous torque measurements. The data is given as a simple coupling factor with respect to the applied angular velocity (as predicted by superconducting-like models) as well as an enhancement factor amplifying the classical frame-dragging field of the spinning source. We see that the effect scales well with an enhancement factor of $\cong 1.10^{18}$ throughout the different experimental setups. Apart from the parity violation, this suggests that the effect behaves similar to a classical frame-dragging field but greatly amplified. It does not show the signature of a superconductivity-like phenomenon, that would lead to similar $B_g/\omega$ coupling factors between the different setups.

**TABLE 1.** Comparison of All Experimental Data for a Frame-Dragging-Like Field at T=4K at the Center of the Spinning Source.

| Experiment | Material | Coupling Factor $(B_g/\omega) \cdot 10^8$ | | Enhancement Factor $\gamma \cdot 10^{18}$ | |
|---|---|---|---|---|---|
| | | CW | CCW | CW | CCW |
| Tajmar et al (Laser Gyro) | YBCO+Al | $5.3 \pm 0.2$ | $-1.2 \pm 0.1$ | $1.7 \pm 0.1$ | $-0.4 \pm 0.1$ |
| | Nb+Al | $3.2 \pm 0.5$ | $-0.4 \pm 0.3$ | $1.2 \pm 0.2$ | $-0.1 \pm 0.1$ |
| | Al | $3.8 \pm 0.3$ | $-0.7 \pm 0.3$ | $1.7 \pm 0.1$ | $-0.3 \pm 0.1$ |
| Graham et al | Pb | $-5.3 \pm 8.5$ | $37.7 \pm 13.2$ | $-0.5 \pm 0.8$ | $1.9 \pm 0.7$ |
| Gravity Probe-B Upper Limit[*] | $SiO_2$ | < 0.1 | | < 0.6 | |
| | Nb | < 0.1 | | < 500 | |

[*] Average over all gyros

The results can also be used to compare with different theoretical models that have been proposed predicting large frame-dragging fields around rotating superconductors. Apart from the parity violation and the non-superconductor critical temperatures observed in the experiments, especially the Gravity-Probe B data rules out all present models by up to 4 orders of magnitude. The experimental data also rules out our initial Cooper-pair mass anomaly hypothesis (Tajmar and de Matos, 2003, Tajmar and de Matos, 2005) by 5 orders of magnitude.

**TABLE 2.** Comparison of Existing Theoretical Models with Experimental Limits at T=4K at the Center of the Spinning Source.

| Theory | Coupling Factor $(B_g/\omega) \cdot 10^8$ | | |
|---|---|---|---|
| | Tajmar Config (Nb) | Graham Config | Gravity Probe-B Config |
| Tajmar and de Matos | 395 | 332 | 395 |
| Dröscher and Hauser | 130 | 130 | 130 |
| de Matos and Beck | 1.6 | 1.6 | 1.6 |
| Experimental Results[*] | $3.2 \pm 0.5$ | $37.7 \pm 13.2$ | < 0.1 |

[*] We neglect the parity violation and use the maximum value from the CW or CCW direction

The gyro responses do not correlate with the accelerometer measurements because they are lower by a factor of 100 if one assumes the standard induction law. It is not clear at the moment if systematic effects such as vibration rectification are responsible for the accelerometer mismatch or if this discrepancy between accelerometer- and gyro measurements is correct as this was recently theoretically predicted (de Matos and Beck, 2007).

All laser gyro systematic effects modeled and analyzed so far show that facility artifacts from mechanical friction, magnetic fields or vibration effects are at least 3 orders of magnitude below the high-resolution gyro measurements. Although vibration offsets might still be present in our data, which will be investigated in further testing, all data and analysis suggests that the observed effects are real.

# NOMENCLATURE

- $A$ = surface area of sample holder (m$^2$)
- $B_g$ = frame-dragging field (rad.s$^{-1}$)
- $c$ = speed of light (= $3 \times 10^8$ m.s$^{-1}$)
- $\gamma$ = enhancement or relative frame-dragging permeability factor
- $G$ = gravitational constant (= $6.67 \times 10^{-11}$ m$^3$.kg$^{-1}$.s$^{-2}$)
- $g$ = gravitational field (in unit of Earth standard acceleration = 9.81 m.s$^{-2}$)
- $I$ = angular momentum (kg.m$^2$.s$^{-1}$)
- $m$ = mass (kg)
- $\eta$ = viscosity (Pa.s)
- $\dot{\omega}$ = angular acceleration (rad.s$^{-2}$)
- $\omega$ = angular velocity (rad.s$^{-1}$)
- $\Omega$ = gyro precession (rad.s$^{-1}$)
- $\psi$ = gyro misalignment angle (deg)
- $R$ = radius (m)
- $T$ = temperature (K)
- $T_c$ = critical superconducting temperature (K)
- $v$ = velocity of helium gas (m.s$^{-1}$)
- $\Delta x$ = gap between sample holder and sensor chamber (m)
- $z$ = distance along central spinning axis (m)

# ACKNOWLEDGMENTS

This research program is funded by the Austrian Research Centers GmbH – ARC. Part of this research was sponsored by the European Space Agency under GSP Contract 17890/03/F/KE and by the Air Force Office of Scientific Research, Air Force Material Command, USAF, under grant number FA8655-03-1-3075. The U.S. Government is authorized to reproduce and distribute reprints for Governmental purposes notwithstanding any copyright notation thereon. We would like to thank R.Y. Chiao, R. Packard, C.J. de Matos and J. Overduin for many stimulating discussions.